\documentclass[superscriptaddress,reprint]{revtex4-1}

\usepackage{subfigure}
\usepackage{graphicx}
\usepackage{natbib}
\usepackage{amsmath}
\usepackage{amssymb}

\begin{document}

%%%%%%%%%%%%%%%%%% title page information %%%%%%%%%%%%%%%%%%
\title{Fresnel reflection from a cavity with net roundtrip gain}

\author{Tobias S. Mansuripur}
\email{mansuripur@physics.harvard.edu} 
\affiliation{Department of Physics, Harvard University, Cambridge, MA 02139}
\author{Masud Mansuripur}
\affiliation{College of Optical Sciences, The University of Arizona, Tucson, AZ 85721}

\begin{abstract}
A planewave incident on an active etalon with net roundtrip gain may be expected to diverge in field amplitude, yet Maxwell's equations admit only a convergent solution. By examining a Gaussian beam obliquely incident on such a cavity, we find that the ``side-tail'' of the beam leaks into the cavity and gives rise to a field that interferes with the main portion of the beam, which is ultimately responsible for the convergence of the field. This mechanism offers perspective for many phenomena, and we specifically discuss the implications for amplified total internal reflection.
\end{abstract}

\maketitle
%PRL abstract: 600 characters

%%%%%%%%%%%%%%%%%%%%%%%%%%  body  %%%%%%%%%%%%%%%%%%%%%%%%%%
The Fresnel coefficients govern the reflection and transmission of light for the simplest possible scenarios: at planar interfaces between homogeneous media. Despite this simplicity, some interesting solutions have been discovered only recently, such as 1) the amplification of evanescent waves in a passive, negative-index slab \cite{pendry_PRL2000,pendry_APL2003,garcia_prl2002}, and controversy regarding the proper choice of the wavevector in active media has persisted in relation to the possibility of 2) negative refraction in nonmagnetic media \cite{chen_PRL2005,ramakrishna_PRL2007,ramakrishna_OptLett2005,geddes_OptComm2007,nistad_PRE2008} as well as 3) single-surface amplified total internal reflection (TIR) \cite{koester_IEEE1966,callary_OptSciAm1976,fan_OptExp2003,willis_OptExp2008,siegman_OPN2010,grepstad_OptExp2011}. It turns out that all three of these cases share a common thread: the presence of a cavity whose roundtrip gain exceeds the loss. In this Letter, we explore more generally the response of such a cavity to an incident beam of light.

\begin{figure}
\includegraphics[scale=0.5]{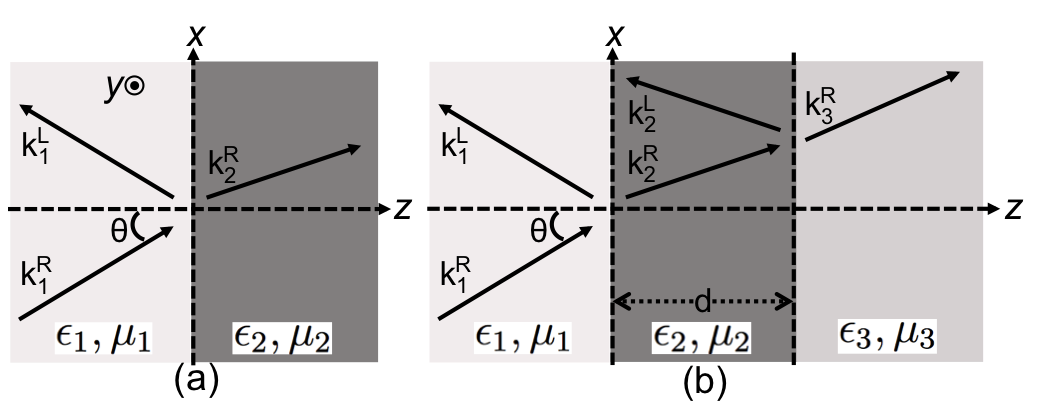}
\caption{\label{fig:geometry}Geometry of the (a) single-surface  and (b) cavity problems.  All media are infinite in the $x$ and $y$-directions. The arrows denote the wavevectors of the planewaves present in each layer. The material constants are the relative permittivities and permeabilities.}
%In the cavity problem, a slab of thickness $d$ with material constants ($\epsilon_2$,$\mu_2$) is placed between two semi-infinite media with constants ($\epsilon_1$,$\mu_1$) and ($\epsilon_3$,$\mu_3$).}
\end{figure}

To begin, we establish conventions that allow us to more clearly discuss the direction of energy flow, which is central to our overall argument. For the single-surface problem, shown in Fig.\ \ref{fig:geometry}(a), the incident wavevector in medium one is $\boldsymbol{k}_1^R = k_x \boldsymbol{\hat{x}} + k_{1z}^R \boldsymbol{\hat{z}}$, and the reflected wavevector is  $\boldsymbol{k}_1^L = k_x \boldsymbol{\hat{x}} + k_{1z}^L \boldsymbol{\hat{z}}$, where $k_{1z}^L = -k_{1z}^R$. The component $k_x$ (which we assume for simplicity to be real-valued), once determined by the incident wave, must become the $x$-component of every wavevector in the system in order to satisfy Maxwell's boundary conditions. For the transmitted wavevector $\boldsymbol{k}_2$, the dispersion relation offers two choices for $k_{2z}$,
\begin{equation}
k_{2z} = \pm \sqrt{(\omega/c)^2 \mu_2 \epsilon_2 - k_x^2} \label{eq:k2zdispersionrelation},
\end{equation}
where $\omega$ is the angular frequency of the planewave and $c$ is the speed of light in vacuum. We denote by $k_{2z}^R$ ($k_{2z}^L$) the choice which carries energy to the right (left), namely, that for which the time-averaged $z$-component of the Poynting vector is positive (negative). (In cases where both choices for $k_{2z}$ result in no energy flow in the $z$-direction, such as for evanescent waves in a transparent medium, our prescription is to add a small amount of loss to the slab which will unambiguously distinguish $k_{2z}^R$ and $k_{2z}^L$, then take the limit as the loss goes to zero. See the supplemental information for details.) Let us postulate that the proper choice for $k_{2z}$ in the single-surface problem is always $k_{2z}^R$ (i.e., that the transmitted energy flows away from the interface), irrespective of the material parameters or the nature of the incident wave. (In fact, $k_{2z}^R$ is universally agreed to be the correct choice in all cases except possibly that of amplified TIR; it is due to this one controversy that we refer to this choice as a postulate for now.) We require this postulate in order to unambiguously define the single-surface Fresnel reflection and transmission coefficients
\begin{equation}
r_{\ell m} = \frac{\tilde{k}_{\ell z}^R - \tilde{k}_{m z}^R}{\tilde{k}_{\ell z}^R + \tilde{k}_{m z}^R},\ t_{\ell m} =\frac{2 \tilde{k}_{\ell z}^R}{\tilde{k}_{\ell z}^R + \tilde{k}_{m z}^R}  \label{eq:ssr}
\end{equation}
where we have generalized the result for incidence medium $\ell$ and transmission medium $m$. For $s$-polarization we have defined $\tilde{k}_{n z} \equiv k_{n z}/\mu_n$, and the two coefficients yield the reflected and transmitted amplitudes of the component $E_y$, while for $p$-polarization $\tilde{k}_{n z} \equiv k_{n z}/\epsilon_n$ and the coefficients are associated with the component $E_x$.

Having established these conventions, we now consider the case of light incident on a cavity, shown in Fig.\ \ref{fig:geometry}(b). The total $E$-field resulting from an $s$-polarized incident wave in medium one with amplitude $E_1^R$ is given by
\begin{equation} \label{eq:Efield}
E_y(x,z) =
\left\{
\begin{array}{ll}
       E_1^R \exp(i k_x  x + i k_{1z}^R z)  &\\
       \ \ \ + E_1^L \exp(i k_x  x + i k_{1z}^L z) & : z \leq 0 \\
       E_2^R \exp(i k_x  x + i k_{2z}^R z) &\\
       \ \ \ + E_2^L \exp(i k_x  x + i k_{2z}^L z) & : 0 \leq z \leq d \\
       E_3^R \exp[i k_x  x + i k_{3z}^R (z-d)] & : z \geq d 
     \end{array}
\right.
\end{equation}
where  $E_1^L$ is the reflected wave amplitude, $E_2^R$ and $E_2^L$ correspond to the two counter-propagating waves in medium two, $E_3^R$ is the transmitted wave amplitude, and the time-dependence $\exp(-i \omega t)$ has been omitted.  We confine our attention to situations where medium three is passive, so that $k_{3z}^R$ is uncontroversially the correct choice for the wavevector in medium three. The most direct route to solve for the four unknown wave amplitudes is to enforce Maxwell's boundary conditions: the transverse components of the $E$ and $H$-fields must be continuous across the two interfaces at $z=0$ and $z=d$, which results in four equations that can be solved for the four unknowns. The resulting reflection coefficient from the slab can be expressed in terms of the single-surface Fresnel coefficients as
\begin{equation}
r \equiv \frac{E_1^L}{E_1^R} = \frac{r_{12} + r_{23} \exp(2i k_{2z}^R d)}{1-\nu} \label{eq:refcoeff}
\end{equation}
where
\begin{equation}
\nu = r_{21} r_{23} \exp(2 i k_{2z}^R d) \label{eq:nu}
\end{equation}
is referred to as the roundtrip coefficient; the amplitude of a planewave circulating in the slab is multiplied by this factor after each roundtrip in the absence of any sources outside the slab. (Although we explicitly discuss $s$-polarized light, our conclusions as well as Eqs.\ \ref{eq:refcoeff} and \ref{eq:nu} hold for both polarization states.) It is essential to note that the reflection coefficient given by Eq.\ \ref{eq:refcoeff} is invariant under the transformation $k_{2z}^R \leftrightarrows k_{2z}^L$; this is not surprising as it can be interpreted simply as a relabeling of the waves $E_2^R \leftrightarrows E_2^L$  in Eq.\ \ref{eq:Efield} that does not affect the final result. This invariance is important because it means that Eq.\ \ref{eq:refcoeff}--which gives the reflection from the slab--is correct even if our postulate about the correct choice for $k_{2z}$ in the single-surface problem turns out to be incorrect. We emphasize that the reflection coefficient given by Eq.\ \ref{eq:refcoeff} is a valid solution to Maxwell's equations for all material parameters, and in particular for any value of $\nu$. The roundtrip coefficient $\nu$ has an important physical meaning, and intuitively one would expect three different regimes of behavior when the magnitude of $\nu$ is less than, equal to, or greater than one. The case where $|\nu|<1$ governs passive slabs (in most but not all cases) and sufficiently weakly amplifying slabs. When $\nu=1$ the slab behaves as a laser and emits light even in the absence of an incident wave, which manifests itself mathematically as an infinitely large reflection amplitude. The case where $|\nu|>1$, however, has received scant attention in the literature \cite{callary_OptSciAm1976}.

Perhaps the reason for the neglect of the $|\nu|>1$ steady-state solution is the seemingly intuitive assumption that there cannot be a steady-state solution when $|\nu|>1$ (due to gain saturation in a laser, for instance. See the supplemental information for details.) This assumption is only reinforced by examining a second well-known solution method for the reflection coefficient that decomposes the reflected wave amplitude $E_1^L$ into a sum over partial waves, yielding the reflection coefficient
\begin{equation} \label{eq:rcoeffsummethod}
r = r_{12} + t_{12}t_{21}r_{23} \exp(2 i k_{2z}^R d) \sum_{m=0}^\infty \nu^m.
\end{equation}
Heuristically, the first term $r_{12}$ (hereinafter referred to as the ``specular'' partial wave) of Eq.\ \ref{eq:rcoeffsummethod} results from the single-surface reflection of the incident wave at the 1-2 interface, and the geometric series accounts for the contributions to the reflected wave following multiple roundtrips within the slab. When $|\nu|<1$, the geometric series in Eq.\ \ref{eq:rcoeffsummethod} converges to $(1-\nu)^{-1}$, giving the same result as found by matching the boundary conditions in Eq.\ \ref{eq:refcoeff}. When $|\nu|>1$, however, the geometric series diverges and the reflection coefficient is infinite. Intuitively, this divergence seems reasonable, since we expect any light that couples into a slab with $|\nu|>1$ to be amplified after each roundtrip, and therefore grow without bound. Nevertheless, Eq.\ \ref{eq:refcoeff} yields a finite reflection coefficient even when $|\nu|>1$, so how can we reconcile these two very different solutions?

\begin{figure}[t]
\centering
\includegraphics[scale=0.23]{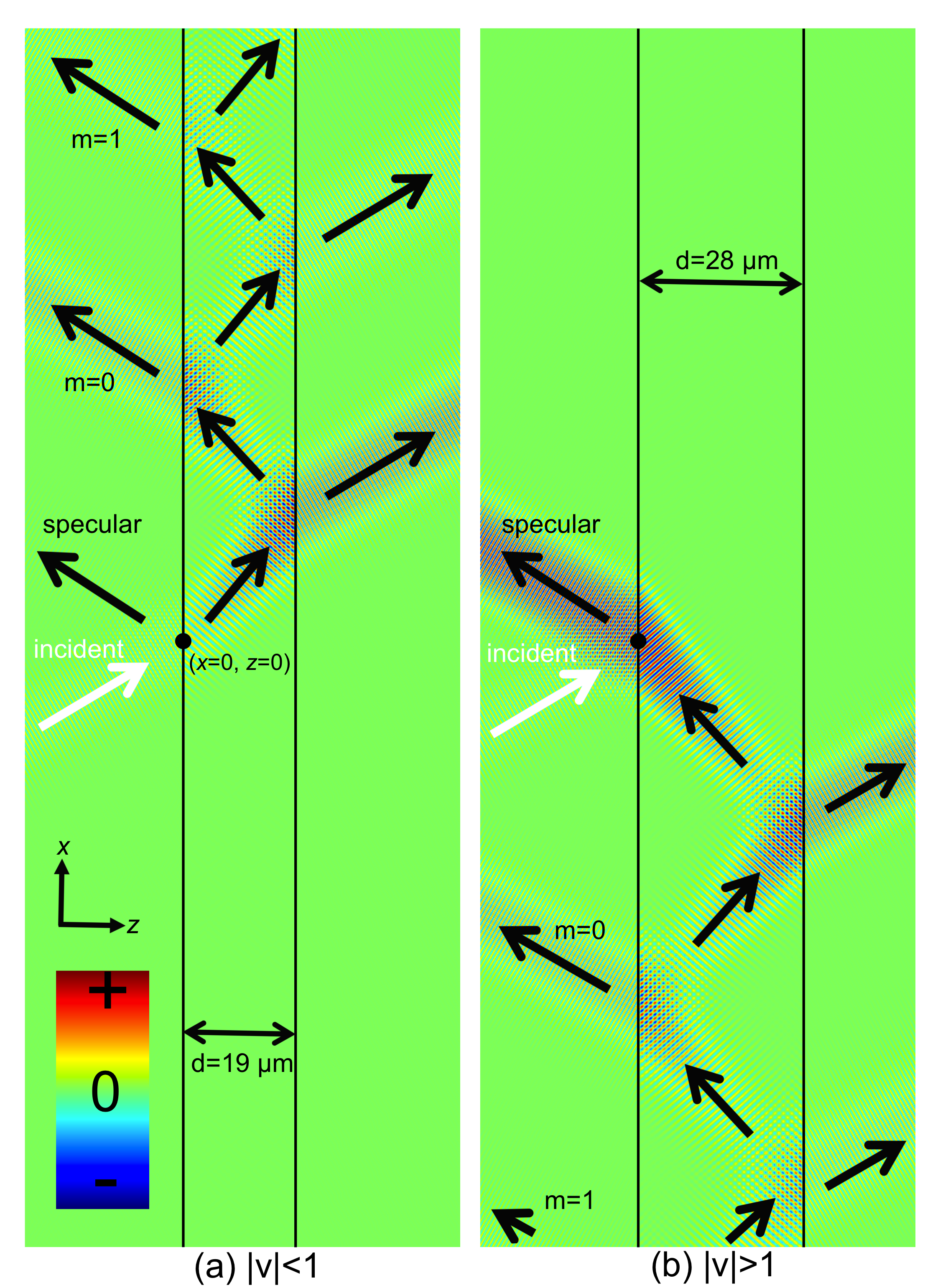}
\caption{\label{fig:simulations}Plots of the field $E_y(x,z)$ at one instant of time for a Gaussian beam (indicated with the white arrow) incident on an amplifying slab for (a) $d=19$ $\mu$m and (b) $d=28$ $\mu$m. Each reflected beam can be associated with a term in the appropriate partial wave sum, either Eq. \ref{eq:rcoeffsummethod} for (a) or  Eq. \ref{eq:rcoeffsummethodprime} for (b). The black dot indicates the origin of the coordinate system.}
\end{figure}

To understand the non-divergent solution, we examine the behavior of a ``finite-diameter'' beam of light incident on the slab by numerically superposing  the planewave solutions of Eq.\ \ref{eq:Efield}, where $E_1^L$, $E_2^R$, $E_2^L$, and $E_3^R$ are all determined by the convergent method of matching the boundary conditions (see supplementary information for details). Let us consider the case where $\epsilon_1 = \epsilon_3 = 2.25$, the slab is an amplifying medium with $\epsilon_2 = 1 - 0.01i$, and $\mu_1  = \mu_2 = \mu_3 = 1$. We superpose a finite number of planewaves with incident angles in the range $27.47^\circ<\theta<32.53^\circ$ and with amplitudes appropriate to generate a Gaussian (to within the sampling accuracy) beam incident on the slab at $30^\circ$ with a full-width at half-maximum beam-diameter of 13.3 $\mu$m. The free-space wavelength of the beam is $\lambda_{\rm o} = 1$ $\mu$m. We can examine the transition at $|\nu|=1$ simply by varying $d$, since both $|r_{21}|$ and $|r_{23}|$ are less than one (and independent of $d$), whereas $|\exp(2ik_{2z}^R d)|$ (and hence $\nu$) increases monotonically with $d$ (because $k_{2z}^R$ has a negative imaginary part). A plot of the field $E_y(x,z)$ at one instant of time is shown in Fig.\ \ref{fig:simulations}(a) for $d=19$ $\mu$m, which was chosen so that $|\nu|$ is slightly less than one for all constituent planewaves of the beam ($0.46<|\nu|<0.99$). The arrows overlying the plot point in the direction of the time-averaged Poynting vector within their vicinity, indicating the direction of energy flow in the system, and the incident beam is uniquely identified by the white arrow. The beam behaves as we expect it to: the incident beam strikes the slab near ($x=0$, $z=0$), giving rise to a specularly reflected beam as well as a refracted beam that `zig-zags' up the slab, which in turn generates a reflected beam in medium one each time it strikes the 2-1 interface. (The field amplitude is plotted on a linear scale, and so the incident beam as well as the specularly reflected beam appear faint relative to the subsequently amplified portions of the beam.) Each of these reflected beams can intuitively be associated with a term of the partial wave expansion of Eq. \ref{eq:rcoeffsummethod}--either the specular term or the $m$th term of the geometric series.

In Fig.\ \ref{fig:simulations}(b) all parameters are kept the same except the slab thickness is increased to $d=28$ $\mu$m, resulting in $|\nu|$ greater than one for all constituent planewaves of the Gaussian beam ($1.01<|\nu|<2.58$). Based solely on the plot of the field amplitude and not on the direction of energy flow indicated by the arrows, it may appear that the incident beam strikes the interface and negatively refracts in the slab, then zig-zags downwards in the $-\boldsymbol{\hat{x}}$ direction, giving rise to many reflected beams in medium one (and transmitted beams in medium three) which emanate from points on the slab with $x<0$. Such an explanation was offered for simulations similar to ours \cite{geddes_OptComm2007,nistad_PRE2008} to attempt to justify negative refraction in an active, nonmagnetic medium. However, by analyzing the Poynting vector we see that the energy in the beam zig-zags {\em up} the slab, so this phenomenon is distinct from negative refraction, despite the similarity in the positions of the reflected and transmitted beams. (In the supplemental information, a video of the time-dependent behavior of a ``finite-duration'' pulse of light more clearly illustrates the direction of energy flow.) The presence of energy in the slab at $x < 0$ has a perfectly causal explanation when one considers that the Gaussian beam does not have a truly finite spatial width, but rather a rapidly decaying ``side-tail'' in the direction normal to the propagation direction. The side-tail is capable of injecting a small amount of energy into the slab at positions $x \ll 0$. When $|\nu|>1$,  light in the slab gains more during one roundtrip than it loses to transmission at both facets, and so this initially small amount of energy is amplified, resulting in the ``pre-excited" field seen at $x<0$ in Fig.\ \ref{fig:simulations}(b), so-called because the excitation occurs before the central lobe of the incident beam arrives at the slab. The key point is that when $|\nu|>1$, our intuition about the arrival time and arrival position of the beam (or pulse) misleads us because amplification by the slab acts on typically negligible field amplitudes to dramatically alter the character of the field. Importantly, we see in Fig.\ \ref{fig:simulations}(b) that when the pre-excited beam meets the incident beam at ($x=0$, $z=0$), the interference is such that all the energy in the slab leaves with the specularly reflected beam. In hindsight, this is necessary for the field to not diverge, since any energy remaining in the slab at this point would continue to zig-zag up and grow without bound. Finally, we emphasize that the specularly reflected beam is amplified relative to the incident beam as a result of the energy it receives from the pre-excited field in the slab, a mechanism that does not occur when $|\nu|<1$.

Although the partial wave method predicts a divergent reflection coefficient when $|\nu|>1$, with one small modification this method in fact offers significant insight into the $|\nu|>1$ case. Recall that the reflection coefficient of Eq.\ \ref{eq:refcoeff} is invariant under the exchange $k_{2z}^R \leftrightarrows k_{2z}^L$.  Applying this same transformation to the partial wave sum of Eq. \ref{eq:rcoeffsummethod} \cite{callary_OptSciAm1976}, we can express the reflection coefficient as
\begin{equation} \label{eq:rcoeffsummethodprime}
r = r_{12}' + t_{12}' t_{21}' r_{23}'  \exp(2 i k_{2z}^L d) \sum_{m=0}^\infty {\nu'}^m,
\end{equation}
where the prime indicates the substitution $k_{2z}^R \rightarrow k_{2z}^L$. Because the new roundtrip coefficient, $\nu' = r_{21}' r_{23}' \exp(2 i k_{2z}^L d)$, is equal to $\nu^{-1}$, in cases where $|\nu|>1$ the primed partial wave sum of Eq. \ref{eq:rcoeffsummethodprime} will converge to the reflection coefficient of Eq. \ref{eq:refcoeff}. Therefore, each reflected beam in Fig.\ \ref{fig:simulations}(b) can be associated either with the specular term $r_{12}'$ or with the $m$th term of the primed partial wave expansion in Eq. \ref{eq:rcoeffsummethodprime}. In particular, note that the amplitude of the specularly reflected beam is $r_{12}$ when $|\nu|<1$, which discontinuously changes to $r_{12}'$ when $|\nu|>1$. Because $|r_{12}|<1$ (in most cases of practical interest) and $r_{12}' = r_{12}^{-1}$, this is a mathematical explanation for why the specular beam is amplified relative to the incident beam only when $|\nu|>1$. Physically, we have seen from Fig.\ \ref{fig:simulations}(b) that this specular amplification is made possible by the pre-excitation, a mechanism which cannot occur when $|\nu|<1$.

It is interesting to compare a lossy and an amplifying slab in the limit as $d \rightarrow \infty$. In a lossy slab (for which ${\rm Im}(k_{2z}^R)>0$), the roundtrip coefficient $\nu \rightarrow 0$ as $d\rightarrow \infty$, and so the reflection coefficient $r$ approaches the single-surface solution $r_{12}$, as expected, because the geometric series in Eq.\ \ref{eq:rcoeffsummethod} makes no contribution. In a gainy slab (for which ${\rm Im}(k_{2z}^R)<0$), $\nu \rightarrow \infty$ as $d\rightarrow \infty$, but $\nu' \rightarrow 0$ and so we see from Eq.\ \ref{eq:rcoeffsummethodprime} that $r \rightarrow r_{12}'$ (which means that the field in the slab is dominated by the wavevector $k_{2z}^L$, i.e., $E_2^R/E_2^L \rightarrow 0$). The reason the limiting treatment of $d\rightarrow \infty$ for a gainy slab does not yield the proper single-surface reflection coefficient is that no matter how large one chooses to make $d$, the nonzero reflection $r_{23}$ at the back-facet of the slab allows for the amplification of the pre-excited field; for $|\nu| \gg 1$, this results in the left-propagating wavevector $k_{2z}^L$ dominating the behavior of the slab while the amplitude of the wave associated with $k_{2z}^R$ diminishes substantially, and the reflection coefficient correspondingly approaches $r_{12}'$. Nevertheless, the right-propagating wave is essential in spite of its seemingly inconsequential amplitude, as it is responsible for generating the left-propagating wave by way of the back-facet reflection. In the case of two truly semi-infinite media (i.e., media one and two), the absence of a back-facet prevents any roundtrip amplification of the pre-excitation, so the only wavevector that exists in the transmission medium is $k_{2z}^R$ and the single-surface reflection coefficient is correctly given by $r_{12}$, not $r_{12}'$.

So far we have examined the relevance of the roundtrip coefficient $\nu$ only through its monotonic dependence on $d$, but $\nu$ is also a function of the incidence angle $\theta$. For the same parameters as those used in Fig.\ \ref{fig:simulations}(b), $\nu$ increases monotonically with an increasing incidence angle $\theta$ for $s$-polarized light; in particular, $|\nu|$ exceeds one as long as $\theta>27.43^\circ$. (For $p$-polarized light, $\nu$ increases monotonically with $\theta$ only once $\theta_2$, the angle of propagation in medium two, exceeds the Brewster angles at both the 2-1 and 2-3 interfaces). As $\theta$ approaches and surpasses the critical angle for TIR, $\theta_c=41.8^\circ$, $|\nu|$ quickly becomes extremely large due to the negatively increasing ${\rm Im}(k_{2z}^R)$. (For $\theta=41^\circ$, $|\nu|= 9.34\cdot 10^3$, and for $\theta=42^\circ$, $|\nu|= 1.40\cdot 10^{15}$.) Thus, TIR from a gainy slab is well within the regime $|\nu|>>1$ (for any reasonable thickness $d$), which, as previously argued, results in a reflection coefficient $r_{12}'$, and therefore the specular beam is amplified. It has been argued extensively that such amplification of the reflected beam is also possible when the gainy medium is truly semi-infinite \cite{koester_IEEE1966,callary_OptSciAm1976,fan_OptExp2003,willis_OptExp2008,grepstad_OptExp2011}; in other words, that the incident wave directly excites the wave with wavevector $k_{2z}^L$ in medium two, resulting in the single-surface reflection coefficient $r_{12}'$. (This conjecture is known as single-surface amplified TIR.) It seems to us that a more unified and consistent approach would be to understand the situation $\theta>\theta_c$ simply as one way to achieve very large $|\nu|$ in a cavity. This would then be comparable to the case of large $d$, for which we demonstrated in the previous paragraph that the existence of the left-propagating $k_{2z}^L$ relies on the nonzero back-facet reflection $r_{23}$ \cite{siegman_OPN2010}. This suggests that $k_{2z}^R$ is the correct choice for the transmitted wavevector in the single-surface problem, even in the case of TIR from an amplifying medium.

For potential future research directions into the pre-excitation mechanism and its consequences, see the supplementary information.

The pulse simulation was run on the Odyssey cluster supported by the Harvard FAS Research Computing Group. TSM is supported by an NSF Graduate Research Fellowship. We thank Alexey Belyanin for helpful discussions.

\end{document}

\title{Fresnel reflection from a cavity with net roundtrip gain: Supplementary Information}

\author{Tobias S. Mansuripur}
\email{mansuripur@physics.harvard.edu} 
\affiliation{Department of Physics, Harvard University, Cambridge, MA 02139}

\author{Masud Mansuripur}
\affiliation{College of Optical Sciences, The University of Arizona, Tucson, AZ 85721}

\maketitle

\section{Prescription for R and L superscripts} 
The energy flux of an $s$-polarized planewave whose $E$-field is given by
\begin{equation}
E_y(x,z,t) = E_0 \exp(i k_x  x + i k_z z - i \omega t)
\end{equation}
in a medium ($\epsilon$, $\mu$) is given by the time-average of the Poynting vector $\boldsymbol{S} \equiv \vec{E} \times \vec{H}$,
\begin{equation} \label{eq:Poyntingvector}
\langle \vec{S} \rangle = \frac{|E_0|^2}{2\omega \mu_0} e^{-2{\rm Im}(k_z )z} \left( {\rm Re}\left[ \frac{k_x}{\mu} \right]\hat{x} + {\rm Re}\left[ \frac{k_z}{\mu} \right] \hat{z} \right).
\end{equation}
Therefore, energy flows in the $+z$-direction (`to the right,' in our convention) when ${\rm Re} (k_z/\mu)>0$, and we denote the wavevector $k_z$ which satisfies this condition with the superscript R. Any value $k_z$ for which ${\rm Re} (k_z/\mu)<0$ is accordingly labeled with a superscript L. It follows from these definitions that $k_z^R$ must make an acute angle with $\mu$ in the complex plane. (For $p$-polarized light energy flows in the $+z$-direction when ${\rm Re} (k_z/\epsilon)>0$, and so $k_z^R$ makes an acute angle with $\epsilon$ in the complex plane.)

\begin{figure}[t]
\centering
\includegraphics[scale=0.25]{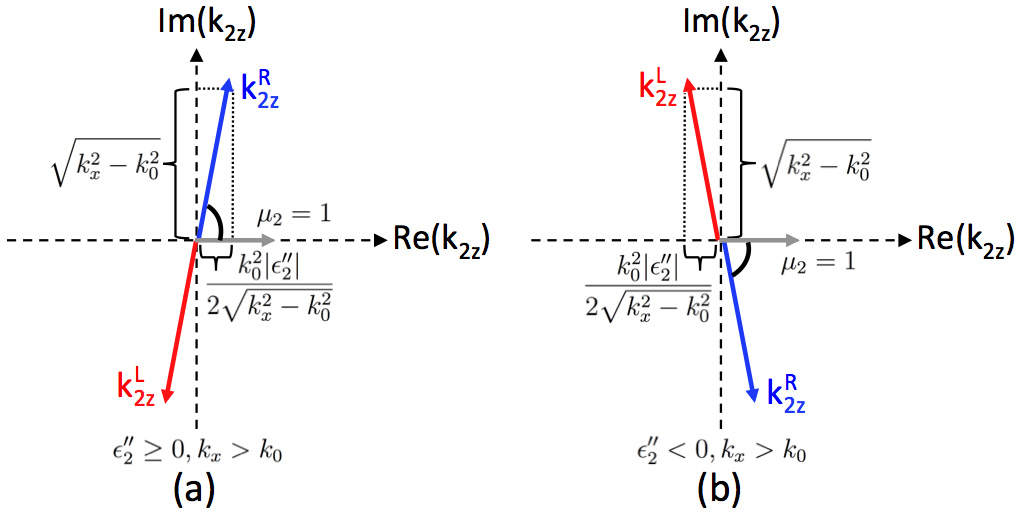}
\caption{\label{fig:RvsLsuperscriptdiagram}Choosing the R or L label for an evanescent wave. (a) The two choices for $k_{2z}$ are shown for the case of a slightly ``lossy vacuum'' ($\epsilon_2 = 1 + i\epsilon_2''$ where $\epsilon_2''>0$, $\mu_2=1$), for the case $k_x>k_0$. The first quadrant solution carries energy to the right and is labeled $k_{2z}^R$, and our prescription is to take the limit $\epsilon_2'' \rightarrow 0$ to determine that $k_{2z}^R$ in the lossless case is along the positive imaginary axis. (b) For a slightly ``gainy vacuum'' ($\epsilon_2'' <0$) the two solutions for $k_{2z}$ are in the second and fourth quadrants, and $k_{2z}^R$ approaches the negative imaginary axis as $\epsilon_2'' \rightarrow 0$. The magnitudes of the real and imaginary parts of $k_{2z}$ in (a) and (b) are approximated using the first order Taylor expansion for small $\epsilon_2''$: $k_0^2 |\epsilon_2''| \ll k_x^2 - k_0^2$.}
\end{figure}

In cases where $\langle S_z \rangle=0$, we must establish a prescription for resolving the ambiguity in the choice of superscript, which is best illustrated by an example. Consider the case where medium one is a lossless dielectric ($\epsilon_1>1$, $\mu_1=1$), medium two is vacuum, and the incident propagating wave satisfies $k_x>k_0$, where $k_0 \equiv \omega/c$, so that the two choices for $k_{2z}$ are $\pm i \sqrt{k_x^2 - k_0^2}$. Both choices for $k_{2z}$ yield pure evanescent waves and carry no energy along the $z$-direction. By adding a small amount of loss to medium two, so that $\epsilon_2 = 1 + i \epsilon_2''$ where $\epsilon_2''>0$, the two choices for $k_{2z}$ deviate slightly from the imaginary axis as shown in Supp.\ Fig.\ \ref{fig:RvsLsuperscriptdiagram}(a). Now both waves carry non-zero energy along the $z$-direction; the first quadrant solution is $k_{2z}^R$ (which can be seen quickly because it makes an acute angle with $\mu_2=1$) and the third quadrant solution is $k_{2z}^L$. Our prescription to establish $k_{2z}^R$ for a true vacuum (i.e., $\epsilon_2''=0$) is to take the limit $\epsilon_2'' \rightarrow 0$, which yields $k_{2z}^R$ as the solution along the positive imaginary axis.

Beware that if one adds a small amount of gain rather than loss to medium two, so that $\epsilon_2 = 1 + i \epsilon_2''$ where $\epsilon_2''<0$, then the two solutions for $k_{2z}$ exist in the second and fourth quadrants as shown in Supp.\ Fig.\ \ref{fig:RvsLsuperscriptdiagram}(b), and in this case $k_{2z}^R$ points predominantly along the {\em negative} imaginary axis. Thus, we see that the two limiting cases as gain or loss approaches zero do not yield the same result:
\begin{equation}
\lim_{\epsilon_2'' \rightarrow 0^+} k_{2z}^R = - \lim_{\epsilon_2'' \rightarrow 0^-} k_{2z}^R.
\end{equation}
To have an unambiguous labeling convention for the case $\epsilon_2''=0$, we emphasize that one must take the limit as {\em loss} approaches zero, which can be different from the limit as gain approaches zero in the case of evanescent waves.

Finally, it is worth noting that this discontinuity in the two limiting cases, apart from being a footnote in establishing a labeling convention, is actually at the heart of the debate over single-surface amplified TIR. When medium two has gain, if one chooses $k_{2z}^R$ as the transmitted wavevector (in accordance with our postulate), then it seems unphysical that as the gain approaches (but does not reach) zero the transmitted wave should still be strongly amplified. To remedy this situation it has been suggested that  the correct choice for the transmitted wavevector should be $k_{2z}^L$ when medium two has gain and $k_x>k_0$, so that the transmitted wave decays in the $+z$-direction. We believe, instead, that the discontinuity in the two limits is not as unphysical as it might appear at first: the transmitted wave propagates a large distance in the $x$-direction while barely moving forward in the $z$-direction (since $k_x \gg {\rm Re}(k_{2z}^R)$), so the large gain in the $z$-direction is actually a result of the long propagation distance along the $x$-direction. Far more unphysical, in our opinion, is the decision to switch the transmitted wavevector from $k_{2z}^R$ when $k_x<k_0$ to $k_{2z}^L$ when $k_x>k_0$. All of these arguments aside, however, the purpose of our paper has been to demonstrate a mechanism by which the specularly reflected beam from a finite-thickness slab can be amplified, both below and above the critical angle.

\section{Gain Saturation}

Physicists familiar with the principles of lasers should be rightfully wary of a steady-state solution with roundtrip coefficient $|\nu|$ greater than one. When an active medium is pumped hard enough to generate a population inversion large enough to yield $|\nu|$ greater than one, light initially generated by spontaneous emission in the cavity will be amplified after each roundtrip. However, the field amplitude does not grow without bound--when the field is large enough, the upper state lifetime is reduced by stimulated emission which causes the population inversion to decrease to a level such that $\nu=1$, resulting in a steady-state lasing solution. This phenomenon of gain reduction with increasing field amplitude is known as gain saturation. In a laser, therefore, the situation $|\nu|>1$ is only a transient state. It's obvious that it cannot be a steady-state solution, because the field would grow without bound.

The situation changes when we allow an incident wave to strike the active medium, as we do in this Letter. Note that $\nu$ is defined as the roundtrip coefficient {\em in the absence of an incident wave}; that is, the reflectivity $r_{21}$ is calculated by assuming that there is no wave in medium 1 arriving at the cavity. We can also define an effective reflectivity $r_{21}^{\rm eff} \equiv E_2^R / E_2^L$ which takes into account the effect of the incident wave. Similarly, we could define an effective roundtrip coefficient in the slab which replaces $r_{21}$ with $r_{21}^{\rm eff}$: $\nu^{\rm eff} = r_{21}^{\rm eff} r_{23} \exp(2i k_{2z}^R d)$. We emphasize that {\em every possible steady-state solution to the problem under consideration, whether the slab is passive or active and whether there is an incident wave or not, satisfies the condition} $\nu^{\rm eff} =1$. This is a property of steady-state solutions: the field in the slab must regenerate itself after every roundtrip, taking into account all sources and sinks. Therefore, in solutions where $|\nu|>1$, the incident wave must interfere destructively with the field in the slab so that $|r_{21}^{\rm eff}| < |r_{21}|$ and ultimately force $\nu^{\rm eff}$ to equal 1. In summary, when there is no incident wave the situation $|\nu|>1$ is temporary because the field will grow until gain saturation (a nonlinear effect) forces the $\nu=1$ solution. With an incident wave, a linear steady-state solution is possible even when $|\nu|>1$ because of the reduction in the effective facet reflectivity $r_{21}^{\rm eff}$, which prevents the unbounded growth of the fields so that we do not have to rely on gain saturation to avoid a nonphysical divergence.

\section{Pulse of light incident on gainy slab with $|\nu|>1$}

The video file pulse\_video.avi included online is a time-lapse video of $|E_y|^2$ of a pulse of light, rather than a beam, for the same material parameters as in Fig.\ 2(b) of the main text: $\epsilon_1 = \epsilon_3=2.25$, $\epsilon_2 = 1 - 0.01i$, $\mu_1=\mu_2=\mu_3 = 1$, $d=28$ $\mu$m. The white vertical lines in the video identify the 1-2 and 2-3 interfaces. The incident pulse is $s$-polarized and Gaussian in both space (FWHM = 13.3 $\mu$m) and time (FWHM = 50 fs, or 15 optical cycles). The central wavelength of the pulse is $\lambda_{\rm o} = 1$ $\mu$m, and the mean incidence angle (i.e., averaged over all constituent planewaves) is $30^\circ$. The size of each video frame is 210 $\mu$m by 150 $\mu$m (height by width). The time elapsed between frames is 10 fs, and the entire video spans 1.22 ps (123 frames total). The field $|E_y|^2$ is plotted on a logarithmic scale covering 3 decades, i.e. red corresponds to the maximum intensity and blue corresponds to intensities less than or equal to 1/1000th of the maximum. The background in this image is blue, which corresponds to the minimum of $|E_y|^2$, whereas the background in Figs. 2(a) and 2(b) is green because it is the field $E_y$ that was plotted in that case, so that blue corresponded to the maximum negative field.

In the video, one first sees the incident pulse near the bottom left of the screen, traveling up and to the right. The pre-excitation is soon seen in the slab at the bottom of the frame, and the reflected pulse that corresponds to the $m=1$ term in the primed partial wave expansion leaves the slab and propagates up and to the left in medium one. The pre-excitation in the slab then undergoes one roundtrip as it zig-zags upward, giving rise to a transmitted pulse in medium three followed by the $m=0$ reflected pulse in medium one. The pre-excitation then makes one more roundtrip, giving rise to another transmitted pulse in medium three, and then approaches the 2-1 interface at the same time the incident pulse arrives from the other side. The two pulses interfere in such a way as to yield an amplified specularly reflected pulse by entirely depleting the energy content of the slab. The fact that the pre-excitation in the slab travels in the $+x$-direction clearly distinguishes this behavior from negative refraction.

\section{Description of Simulation}

The $E$-field plots of the Gaussian beams and the video of the pulse were created using MATLAB. The field at each pixel is determined by superposing a large (but of course finite) number of planewave solutions. In this regard, the plots represent perfectly analytical solutions to Maxwell's equations.

As described in the main text, the response of the slab to an incident $s$-polarized planewave with amplitude $E_1^R$ and wavevector $\boldsymbol{k}_1^R = k_x \boldsymbol{\hat{x}} + k_{1z}^R \boldsymbol{\hat{z}}$ is given by
\begin{equation} \label{eq:Efield}
E_y(x,z) =
\left\{
\begin{array}{ll}
       E_1^R \exp(i k_x  x + i k_{1z}^R z)  + E_1^L \exp(i k_x  x + i k_{1z}^L z) & : z \leq 0 \\
       E_2^R \exp(i k_x  x + i k_{2z}^R z) + E_2^L \exp(i k_x  x + i k_{2z}^L z) & : 0 \leq z \leq d \\
       E_3^R \exp[i k_x  x + i k_{3z}^R (z-d)] & : z \geq d 
     \end{array}
\right.
\end{equation}
and the time-dependence factor $\exp(-i \omega t)$ is not explicitly written. The wavevector components $k_{2z}^R$ and $k_{3z}^R$ are determined by the dispersion relation
\begin{equation}
k_{\ell z}^R = \sqrt{(\omega/c)^2 \mu_\ell \epsilon_\ell - k_x^2} \label{eq:k2zdispersionrelation},
\end{equation}
where $\mu_\ell$ and $\epsilon_\ell$ are the relative magnetic permeability and electric permittivity constants of material $\ell$, and the sign of the square root is chosen according to the prescription described in Supplementary Sec.\ 1. The four unknown wave amplitudes are found by satisfying Maxwell's boundary conditions to be
\begin{align}
E_2^R &= \frac{2 k_{1z}^R (k_{3z}^R + k_{2z}^R)E_1^R}{(k_{2z}^R + k_{1z}^R) (k_{3z}^R + k_{2z}^R) + \exp(2 i k_{2z}^R d) (k_{3z}^R - k_{2z}^R) (k_{2z}^R - k_{1z}^R)} \label{eq:E2R}\\
E_2^L &= \frac{-2 k_{1z}^R (k_{3z}^R - k_{2z}^R)E_1^R}{(k_{2z}^R - k_{1z}^R) (k_{3z}^R - k_{2z}^R) + \exp(-2 i k_{2z}^R d) (k_{3z}^R + k_{2z}^R) (k_{2z}^R + k_{1z}^R)} \\
E_1^L &= E_2^R + E_2^L - E_1^R \\
E_3^R &= E_2^R \exp (i k_{2z}^R d) + E_2^L \exp (-i k_{2z}^R d) \label{eq:E3R}.
\end{align}
To construct the Gaussian beam from the planewave solutions, we begin by expressing $E_y$ in the $z=0$ plane for a  beam traveling parallel to the $z$-axis
\begin{equation}
E_y(x,z=0) = E_0 \exp\left({-\frac{x^2}{2\sigma_x^2}}\right),
\end{equation}
where $E_0$ is the peak amplitude and $\sigma_x$ is directly proportional to the spatial FWHM
\begin{equation}
w_x = 2\sqrt{2 \ln 2} \sigma_x .
\end{equation}
By Fourier transforming and subsequently inverting the transform, the field can equivalently be written as an integral in k-space,
\begin{equation}
\label{eq:fieldintegralz=0}
E_y(x,z=0) =  \int_{-\infty}^{\infty} dk_x E_1^R(k_x) \exp(i k_x x),
\end{equation}
where
\begin{equation}
\label{eq:gaussiankspace}
E_1^R(k_x) = \frac{E_0 \sigma_x}{\sqrt{2\pi}}\exp \left( \frac{-k_x^2}{2(1/\sigma_x)^2} \right),
\end{equation}
and the FWHM in $k$-space is
\begin{equation}
w_k = 2\sqrt{2\ln 2}/\sigma_x .
\end{equation}
To propagate the beam beyond the $z=0$ plane, we associate with each value of $k_x$ a component $k_{1z}^R$ such that the total wavevector obeys the dispersion relation in medium 1, 
\begin{equation}
k_{1z}^R (k_x) = \sqrt{(\omega/c)^2 \mu_1 \epsilon_1 - k_x^2} .
\end{equation}
Now the Gaussian beam can be expressed as a function of $x$ and $z$ by
\begin{equation}
\label{eq:fieldintegral}
E_y(x,z) =  \int_{-\infty}^{\infty} dk_x E_1^R(k_x) \exp[i( k_x x + k_{1z}^R z)].
\end{equation}

At this point, we must approximate the integral in Eq.\ \ref{eq:fieldintegral} by discretization so that the calculation can be carried out by a computer. We restrict $k_x$ to a finite sampling width given by $-w_s/2 <k_x< w_s/2$, and sample the beam equidistantly within this region with a total number of samples $N_s$. The integral in Eq.\ \ref{eq:fieldintegral} is approximated by the sum
\begin{equation}
\label{eq:fieldsum}
E_y(x,z) = \sum_{k_x=-w_s/2}^{w_s/2} \Delta k_x E_1^R (k_x) \exp[i (k_x x + k_{1z}^R z)],
\end{equation}
where
\begin{equation}
\Delta k_x = \frac{w_s}{N_s - 1}.
\end{equation}
At this point, it is helpful to think of $E_1^R$, $k_x$, and $k_{1z}^R$ as vectors containing $N_s$ numerical elements each. To rotate the beam so that it travels at an angle $\theta$ to the $z$-axis, we perform the transformation
\begin{align}
k_x &\rightarrow \cos(\theta) k_x + \sin(\theta)k_z \\
k_{1z}^R &\rightarrow - \sin(\theta) k_x + \cos(\theta)k_{1z}^R
\end{align}
on each element of $k_x$ and $k_{1z}^R$. (The Fourier amplitude of each plane-wave $E_1^R(k_x)$ is unaffected by the rotation in the case of $s$-polarized light.) Finally, to displace the waist of the beam to some location $(x_0,z_0)$ in the incidence medium, one must multiply each Fourier amplitude by
\begin{equation}
E_1^R(k_x) \rightarrow E_1^R(k_x) \exp[-i(k_x x_0 + k_{1z}^R z_0)].
\end{equation}
With these redefined values for $E_1^R$, $k_x$, and $k_{1z}^R$, the sum in Eq. \ref{eq:fieldsum} is a good approximation to a Gaussian beam traveling at an angle $\theta$ whose waist is located at $(x_0,y_0)$. The total $E$-field at any point in the system is given by
\begin{equation}
\label{eq:Efieldinspace}
E_{\rm tot}(x,z) =
\begin{cases}
{\rm Real}\{ \sum \Delta k_x \left( E_1^R (k_x) \exp[i( k_x x + k_{1z}^R z)] + E_1^L(k_x)\exp[i( k_x x + k_{1z}^L z)]\right)\}, & z \leq 0 \\
{\rm Real}\{ \sum \Delta k_x \left( E_2^R (k_x) \exp[i( k_x x + k_{2z}^R z)] + E_2^L(k_x)\exp[i( k_x x + k_{2z}^L z)]\right)\}, & 0\leq z \leq d \\
{\rm Real} \{ \sum \Delta k_x E_3^R (k_x) \exp[i(k_x x + k_{3z}^R z)]\}, & z \geq d
\end{cases}
\end{equation}
where $E_1^L$, $E_2^R$, $E_2^L$, and $E_3^R$ are calculated element-wise from $E_1^R(k_x)$ according to Eqs. \ref{eq:E2R}-\ref{eq:E3R}. The beam plots in Fig.\ 2 of the main text are calculated pixel-by-pixel from the sum in Eq.\ \ref{eq:Efieldinspace}, with the values of $x$ and $z$ indicating the location of the pixel. The resultant field is normalized to the maximum field value in the image, and displayed in color. The pulse video is calculated similarly, except that the field is Gaussian in space and time, and so the field must be sampled in both spatial and temporal frequency. The calculation time is significantly longer for the pulse compared to the beam, and the simulations are only practical on a supercomputer.

The finite nature of the sampling has consequences which must be considered in order to be sure that our results are not affected by numerical artifacts. Firstly, the truncation of the Gaussian beam in $k$-space to the sampling width $w_s$ leads to a convolution with a sinc function in the spatial domain. Therefore, the side-tail of our beam is not truly Gaussian; rather, the envelope of the side-tail is Gaussian but the side-tail itself exhibits periodic sinc-like fluctuations in intensity (which cannot be seen in Fig.\ 2 of the main text, but can be seen in logarithmic plots which resolve the small intensities of the side-tail). The sampling width chosen for Fig.\ 2 was $w_s = 2 w_k$ (with $N_s=501$). We made sure that other choices of the sampling width, $w_s = 3w_k$ and $4w_k$ (with proportionally larger $N_s$ so that $\Delta k_x$ remained constant), did not affect the behavior of the plots. Therefore, our conclusions are not affected by the precise value of the sampling width $w_s$. Secondly, the finite number of samples $N_s$ implies the spectrum of $k_x$ values is discrete, so the incident beam is periodic in space. This means that in the plots of Fig.\ 2 in the main text, there is not just one incident beam but an infinite number of them impinging on the slab, spaced periodically along the $x$-axis by a distance $2\pi / \Delta k_x$ = 2830 $\mu$m. If the sampling is increased from $N_s=501$ to 2001 (while keeping $w_s = 2w_k$ constant), the distance between adjacent beams increases to 11330 $\mu$m, but the plots in both Figs.\ 2(a) and 2(b) look identical to the ones with 501 samples. Therefore, 501 samples is sufficient in this case to ensure the beams do not interfere with each other, and the plot is a good representation of the field of a single beam.

\section{Future work}

Our intent in this Letter has been to demonstrate the unintuitive behavior of a beam incident on a slab whose roundtrip coefficient is greater than one, which we demonstrated using analytical solutions to Maxwell's equations. We argued that when $|\nu|>1$, the amplification of typically negligible field amplitudes results in a ``pre-excited" beam in the slab which interferes with the incident beam to prevent the divergence of the field. We focused on two peculiar consequences of this phenomenon: 1) the specularly reflected beam is amplified only when $|\nu|>1$, and 2) the field in the slab is dominated by the wavevector $k_{2z}^L$ when $|\nu| \gg 1$. Our analysis has been restricted to the case of planar media with infinite extent in the $x$ and $y$-dimensions, with homogeneous and frequency-independent material parameters. We do not consider this a serious drawback of our argument, as the majority of analyses of the three-layer problem employ the same assumptions. From a purely theoretical viewpoint and within the confines of the assumptions we have made, therefore, we believe that these results provide perspective for single-surface amplified TIR, and counter the notion of negative refraction in a nonmagnetic slab. Because the pre-excitation mechanism can only occur in a finite-thickness slab, we speculate that $k_{2z}^R$ is the correct choice for the transmitted wavevector in the single-surface problem in all cases; in other words, the transmitted wave always carries energy away from the interface. There remain many open questions to be answered theoretically. Finite-difference time-domain simulations may be best suited to determine how the pre-excitation mechanism, which begins at $x \ll 0$, is affected when the slab has a finite length in the $x$-direction. Spontaneous emission and gain saturation must be accounted for in real materials. How would the slab behave if the beam had a truly finite width and no side-tail? To investigate how the slab reaches the steady state over time in response to a pulse with a sharp turn-on, the material parameters of the slab must be made to obey the Kramers-Kronig relations. Despite our lack of answers to these important questions, we hope that our analysis has clarified at least some of the relevant issues.

Surprisingly, $|\nu|$ can exceed one even in passive media provided $|r_{21}|$ or $|r_{23}|$ exceeds one, which can happen for incident evanescent waves near surface plasmon resonances. Understanding the role of $\nu$ in these cases can yield additional insight, particularly to the case of Pendry's lens [1, 2], and will be treated in future work.